\documentclass[10pt,a4paper]{article}
\usepackage{graphics}

\newcommand{\bea}{\begin{eqnarray}}
\newcommand{\eea}{\end{eqnarray}}
\newcommand{\be}{\begin{equation}}
\newcommand{\ee}{\end{equation}}

\def\be{\begin{eqnarray}}
\def\ee{\end{eqnarray}}
\def\bd{\begin{displaymath}}
\def\ed{\end{displaymath}}

\def\etal{{\em et al}}
\begin{document}
\title{Neutron rich nuclei in a new binding energy formula and the 
astrophysical $r-$process}
\author{Chirashree Lahiri and G. Gangopadhyay\\
Department of Physics, University of Calcutta\\
92, Acharya Prafulla Chandra Road, Kolkata-700 009, India\\
email: ggphy@caluniv.ac.in}
\maketitle

\begin{abstract}
Neutron rich nuclei has been studied with a new phenomenological
mass formula. Predictions of different mass formulas for the location of the 
neutron dripline are compared with those from the present calculation.
 The implications of the new mass formula for $r$-process 
nucleosynthesis are discussed. It is found that though the neutron drip line
obtained from this formula differs substantially from other formulas,
the $r$-process abundance upto mass 200 are unlikely to be significantly 
different. The errors inherent in the mass formula are found to play an
insignificant role beyond mass $A=80$.
\end{abstract}

\section{Introduction}

Rapid neutron capture or $r$-process nucleosynthesis is an important ingredient 
in the production of heavy elements. This process proceeds through very 
neutron rich regions  of the nuclear landscape inaccessible in terrestrial 
laboratories. It 
is known not to depend strongly on the neutron absorption cross section. On 
the other hand, the ground state binding energy strongly influences the process.
Experimental mass measurements are presently not possible in nuclei in the 
$r$-process path and  one has to take recourse to theoretical predictions. In 
Gangopadhyay\cite{massform} (hereafter called Ref. I), a new 
phenomenological formula for ground state binding energies was introduced. 
In a subsequent work\cite{mass1}, we explored the proton rich side of the 
stability valley and 
studied the implication of the new formula for the rapid proton capture process.

All the phenomenological mass formulas have necessarily been devised by fitting 
the nuclei near the stability valley as the experimental values are available 
only for these nuclei. Neutron rich nuclei are experimentally very difficult to
produce in terrestrial laboratories. It is not possible to test the
performance of the mass formulas for such nuclei. An alternate test of the 
mass formula may be the reproduction of the r-process abundance. 

In the present work we locate the neutron drip line for the mass formula 
of Ref I and compare our results with two standard mass formulas. Next, 
we study the astrophysical $r$-process abundance using the mass formula.
We also investigate the effect that the errors in the predictions may have 
on the $r$-process.

\section{Results}

\subsection{Neutron dripline}

The neutron dripline for a particular neutron number is defined to be the 
nucleus beyond which the neutron separation energy becomes zero or negative. 
Obviously, because of the pairing effect, the heaviest isotope of any element
necessarily will have even neutron number.
Studying the nucleon drip line is one of the major activities in nuclear 
physics. However, it has been possible to reach the neutron dripline in only 
very light nuclei which are beyond the ambit of mass formulas. 
 
\begin{figure}
\center
\resizebox{7.4cm}{!}{ \includegraphics{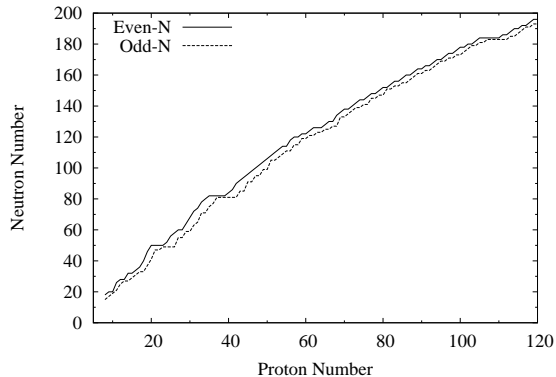}}
\caption{Neutron drip line according to the present calculation.\label{drip}}
\end{figure}

The calculated mass values of nuclei enable us to calculate the location of the
dripline. We determine the location of the neutron dripline from the new 
binding energy formula of Ref. I and present it in Fig. \ref{drip} where we
connect the  neutron number of the last bound isotope, with odd and even number 
of neutrons separately for each element, starting from oxygen. As already noted,
pairing effect ensures that the drip line for even neutron number lies beyond 
that for odd neutron number. 

Gridnev \etal\cite{gridnev} have studied extremely neutron rich Zr and Pb nuclei
in deformed Hartree-Fock approach using various Skyrme forces. They predicted
the existence of stability peninsula in very neutron rich isotopes. They suggested
that it is an effect of the shell structure and 
may be a general feature in neutron rich nuclei. Though our predictions of 
neutron drip line are closer to the stability valley, we find that shell effects are manifested at the drip line also. As magic nuclei are more stable than 
their neighbouring isotopes due to the effect of shell closure, an element with
neutron magic number is more likely to be the last bound nuclei for that 
isotopic chain. It is evident from Fig. \ref{drip} that at $N=50,~82$ and
and 184, the neutron drip line becomes parallel to the $Z$ axis. 
At other shell or subshell closures also, this effect is observed though not 
as prominent as the neutron numbers mentioned above. We also see that at these
three neutron numbers, the drip line for odd and even $N$ come very close 
to each other, indicating the effect of the shell closure. 

One needs to remember that the formula has an average error of 376 keV. Thus, 
if the predicted binding energy of the last neutron in an odd $N$ nucleus is 
very small, it is possible that the nucleus is actually beyond the dripline. 
Conversely, prediction of a very small negative value of last neutron separation
energy cannot guarantee that the dripline does not lie actually beyond. Thus, 
the driplines may conceivably shift by two nucleons either way.

\begin{figure}
\center
\resizebox{7.4cm}{!}{ \includegraphics{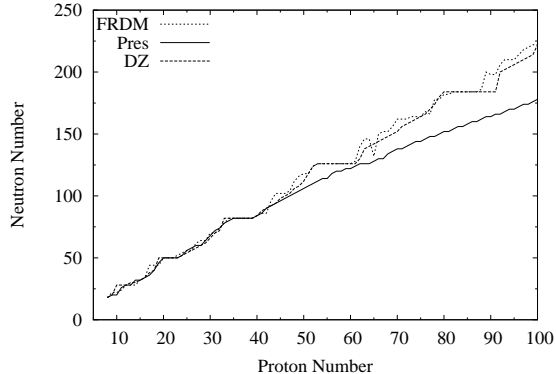}}
\caption{Neutron drip line for even $N$-nuclei predicted by the present formula (Pres), FRDM and Duflo-Zuker (DZ) formula.\label{dripcomp}}
\end{figure}

As evident from Ref I, the neutron drip line predicted by the present formula 
differs substantially from that predicted by other common mass formulas, such as
Finite Range Droplet Model (FRDM)\cite{frdm} or Duflo-Zuker formula\cite{DZ}.
The results for drip line for even neutron nuclei for the present formula are 
compared with the above two calculations in Fig. \ref{dripcomp}. We note that 
beyond $Z\sim 64$, the prediction of the other two formulas differ significantly
from the present calculation. However, in absence of experimental data, it is 
not possible to comment on their relative merits. 

\subsection{Astrophysical r-process}

What about the effect of the new mass formula on nucleosynthesis? 
We have found the present formula to be successful in describing
the astrophysical $rp$-process that occurs in $N\sim Z$ nuclei and 
involves proton rich nuclei\cite{mass1}. Neutron rich nuclei take part in
astrophysical $r$-process where neutrons are captured rapidly by nuclei 
leading to very neutron rich region. These then decay to end up as ordinary
nuclei near the stability valley.

We choose to employ the canonical $r$-process framework within the waiting 
point approximation. In this approach, rapid neutron capture is in equilibrium 
with its inverse process, the ($\gamma,n$) process.  Thus, the 
ratio of the number of nuclei with (Z,N+1) to that of (Z,N) is given by the 
Saha equation.

\be \frac{{\mathcal N}(Z,N+1)}{{\mathcal N}(Z,N)}=
\frac{G^*(Z,N+1)}{2G^*(Z,N)}N_n T^{-3/2}10^{-34.075} 10^{5.04*S_n(Z,N+1)/T}
\ee
where $T$ is the temperature in GK and $G^*=(2J_{gs}+1)G$ are the temperature 
dependent partition functions taken from Rauscher \etal\cite{RT}.

Neutrons are rapidly absorbed by the nuclei reaching very large neutron to 
proton ratio. Beta decay turns the neutrons into protons. Near the  neutron 
shell closures, the $r$-process path approaches the stability valley resulting
in peaks in the abundance distribution. After a certain time interval, 
$r$-process is assumed to stop abruptly. The frozen population then undergoes a 
beta decay cascade, including beta delayed neutron emission, to give the final 
abundance. The details of the method can be found in 
standard text books [see e.g. Iliadis\cite{book}] and reviews\cite{rprev}.

\begin{figure}
\center
\resizebox{7.4cm}{!}{\includegraphics{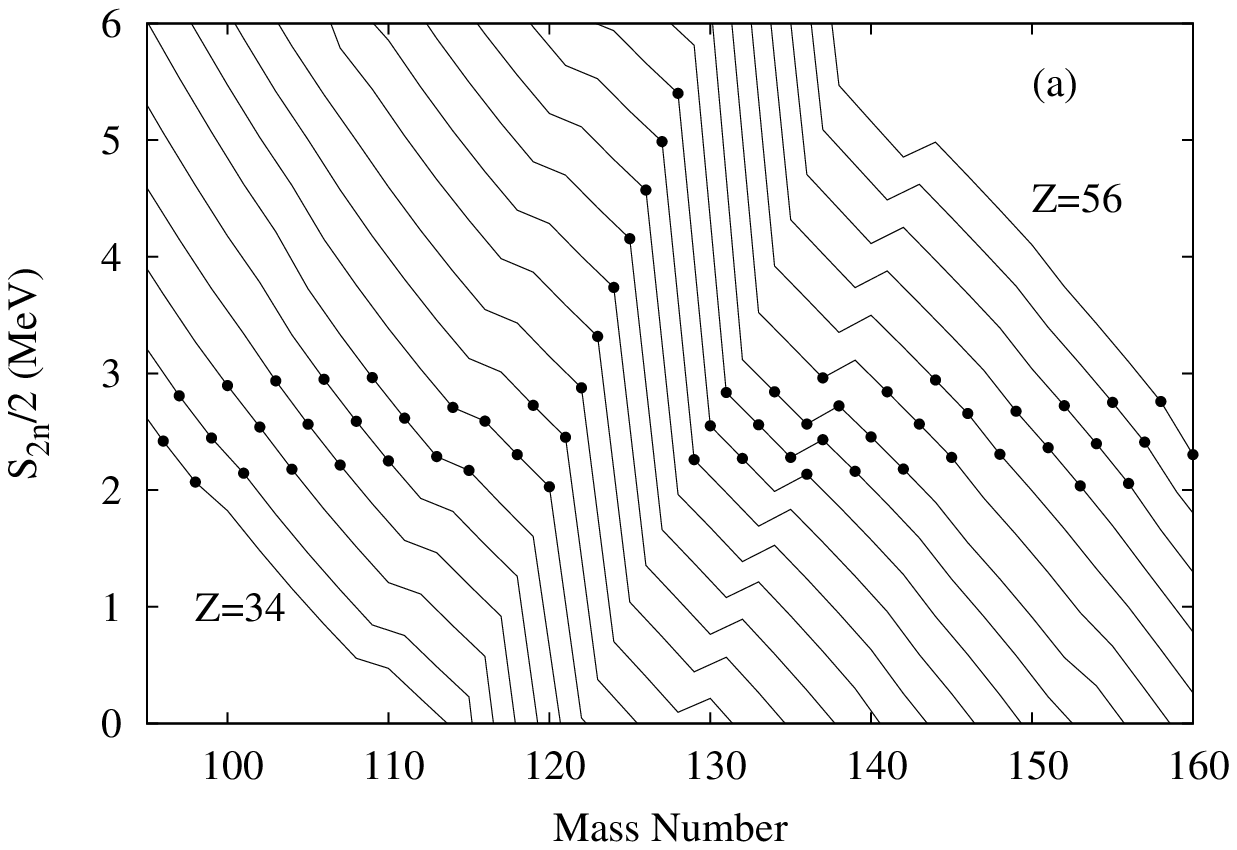}}
\resizebox{7.4cm}{!}{\includegraphics{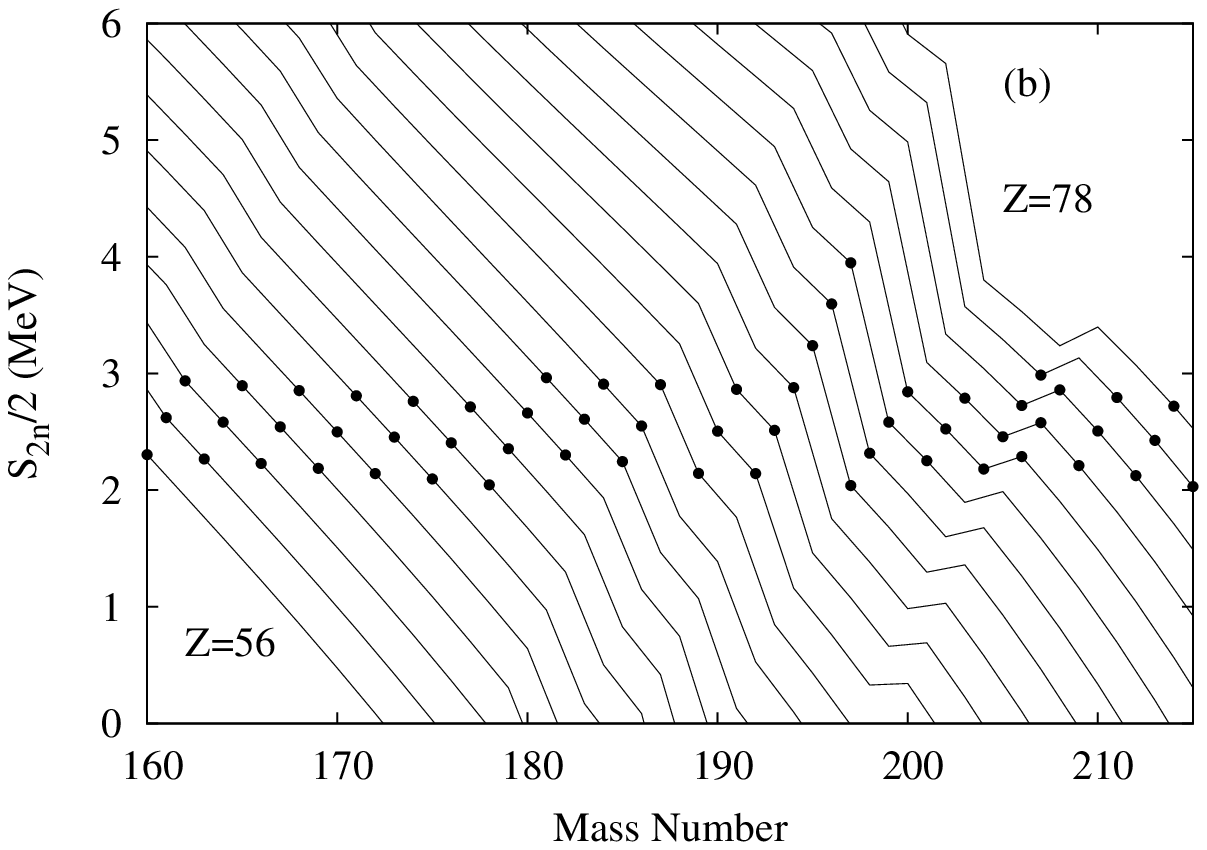}}
\caption{The two neutron separation energy $S_{2n}$ and the $r$-process path. 
\label{path}}
\end{figure}

The typical neutron flux densities densities range from $10^{21}-10^{27}$/cm$^3$
and the temperature varies from 1-2 GK. Putting these
numbers in Saha equation, we find that away from the closed shell, 
the $r$-process proceeds typically along the paths characterised by 
$S_n\approx2-3$ MeV. This signifies that the path lies among neutron rich 
nuclei. In Fig. \ref{path}, we have shown the path followed by the the 
$r$-process. To avoid the odd-even effect, it is a standard procedure to plot
the path in terms of two neutron separation energy $S_{2n}$. The dark circles
indicate the path taken by the $r$-process. When the path nears a closed shell, 
the neutron separation energy increases and the path shifts upwards towards 
more stable nuclei. Thus one expects a peak 
corresponding to the nuclei involved in the the $r$-process with neutron 
number close to a neutron closed shell. In Fig. \ref{path}, we see how the 
shell closures at $N=82$ and $N=126$ turn the $r$-process path towards 
higher proton number {\em i.e.} more stable nuclei.

\begin{figure}
\center
\resizebox{7.4cm}{!}{\includegraphics{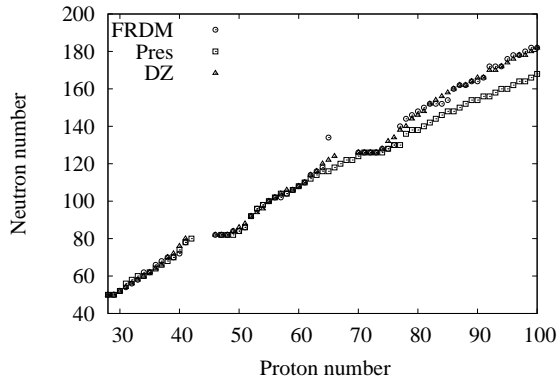}}
\caption{Lightest isotopes for which the two neutron separation 
energy is above 4 MeV. See text for details.\label{range}}
\end{figure}

In view of the difference between the neutron drip line predicted by different 
formula, it should be interesting to look at the differences between the 
$r$-process paths suggested by different mass formulas in Fig. \ref{range}. To keep the 
diagram simpler, we plot the neutron and proton numbers of the lightest 
isotopes of each element between $Z=28$ and 100 for which the two neutron 
separation energy is above 4 MeV. As one can see, upto $A=195$, the results 
of Duflo-Zuker formula and that of Ref I are nearly identical. FRDM predictions 
lie in slightly higher neutron numbers. However, all the three formulas are 
expected to produce  $r$-process abundance peaks at nearly the same mass 
regions.

The abundance curve of $r$-process nuclei show prominent peaks around masses 80,
130 and 195. This corresponds to neutron closed shell $N$= 50, 82 and 126. 
However, it is well known that a single $r$-process cannot explain 
the observed abundance of heavy nuclei including the above three peaks. Some
recent calculations involve a large number of components under various 
conditions of density and temperature to look at the observed abundance. In the 
present work, rather than looking at a large number of possible processes, we 
choose certain standard scenarios and see whether the observed peaks are 
reproduced. 
 
We have calculated the $r$-process abundance in three standard scenarios. 
In all cases, we assume a r-process time of two seconds. The beta decay 
life times and the beta delayed neutron emission probabilities have been taken 
from M\"{o}ller \etal\cite{beta}. We finally add the three results with 
relative weights to compare with the experimental abundance measurements. The 
newly developed formula have been used for masses. The three processes have 
been characterized by the following neutron density and temperature values;
Component I : $N_n=10^{21}/{\rm cm}^3,~~~ T=1.9~GK$, 
Component II : $N_n=10^{25}/{\rm cm}^3,~~~ T=1.5~GK$
and Component III : $N_n=10^{27}/{\rm cm}^3,~~~ T=1.3~GK$. 
The relative weights are taken as 
30, 10 and 1, respectively. The total abundance has been normalized to 
reproduce the abundance at $A=130$. No other attempt has been made to fit the 
experimental data.

\begin{figure}
\center
\resizebox{7.4cm}{!}{\includegraphics{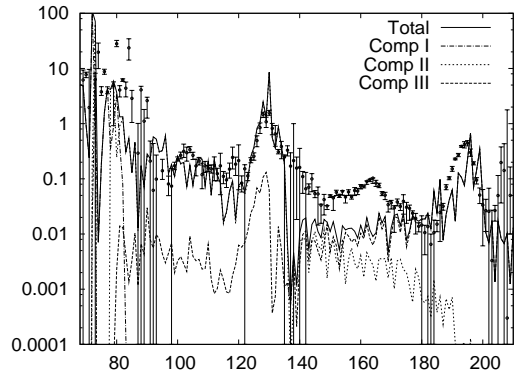}}
\caption{\label{abund}The abundance pattern from a weighted sum of three
$r$-process components. See text for details.}
\end{figure}

Our results are presented in Fig. \ref{abund}. One can see that the peaks 
seen around mass 80, 130 and 195 have been
reproduced to some extent. Usually, one uses a large number of components, with
neutron density, temperature, time and weights fitted to reproduce observed 
data. However, this may actually lead to spurious components which will be 
indistinguishable from an actual event. Rather than fitting the $r$-process
abundance, the aim of the present work is to see whether the peaks in
the abundance distribution can be reproduced using the mass formula of Ref I.

Finally, a comment on the significance of the errors in calculating binding 
energy using the present formula needs to be made. 
Canonical r-process calculations indicate that the nuclei involved in the 
$r$-process have neutron separation energies lying between 2 to 3 MeV. 
As pointed out in Ref I, the root men square (r.m.s.) error 
in the ground state binding energies is 0.376 MeV for 2140 nuclei. 
No global mass formula, be it microscopic or phenomenological,
shows a significantly better agreement. Near the stability valley, single 
nucleon separation energy (proton or neutron) is of the order of 8 MeV. Thus, 
a small error in binding energy does not influence the results of processes 
which involve nuclei near this valley. It is important to investigate the 
effect of the error, that is inherent in all global mass formulas, in the  
determination of the r-process path. 

To see what effect this may have on the final abundance distribution, we have
varied the mass values randomly following a Gaussian distribution with the 
calculated value as mean and standard deviation equal to the rms error quoted 
above. From the distribution of final abundance values for different masses, 
we have extracted the standard deviations. We find that the error plays a 
significant role in light masses. In nuclei beyond $A=80$, inclusion 
of the error does not affect the results significantly.

\section{Summary}

The mass formula developed in Ref. I\cite{massform} has been employed to 
study the nuclei near the neutron drip line. The location of the neutron
drip line has been calculated using the formula. The drip 
line predicted by the present formula differs substantially in heavy nuclei
from predictions for two standard formulas. The abundance pattern resulting 
from the astrophysical $r$-process has been studied in the canonical approach 
under the waiting point approximation. Three different $r$-process scenarios 
have been considered which are seen to give rise to the observed peaks in mass 
80, 130 and 195 region. The differences between the present and the other two  
formulas are unlikely to affect the abundance pattern significantly below 
$A=200$. The error in the mass formula does not significantly affect the 
abundance pattern beyond mass 80.

\section*{Acknowledgments}

This work has been carried out with financial assistance of the UGC sponsored
DRS Programme of the Department of Physics of the University of Calcutta.
Computer facilities created under DST-FIST Programme of the University of 
Calcutta have been utilized. Chirashree Lahiri acknowledges the grant of a fellowship 
awarded by the UGC.

\end{document}